\documentclass[11pt]{article}
\usepackage[margin=2.5cm]{geometry}
\usepackage{amsmath}
\usepackage{amsthm}
\usepackage{amsfonts}

\theoremstyle{definition}

\theoremstyle{remark}
\newtheorem{note}{Note}[section]
\theoremstyle{plain}

\newtheorem{prop}{Proposition}[section]

\newcommand{\calh}{\mathcal{H}}

\begin{document}
\title{Feynman diagrams, Hopf algebras and renormalization}
\author{H.M. Ratsimbarison\\
       Institut @-HEP-Mad, Antananarivo}
\date{December 2005}

\maketitle
\begin{abstract} This paper gives a review of Connes-Kreimer formulation of perturbative renormalization in Quantum Field Theory. We begin with the derivation of the Feynman calculus, the Hopf algebra structure on Feynman diagrams and we show the natural occurence of Birkhoff decomposition in perturbative renormalization. Most of proofs is added in the present paper.
\end{abstract}

\section{Introduction}
In Quantum Field Theory, Feynman graphs are used to calculate correlation functions (or correlators) in the perturbative case. Due to Feynman rules, one can associate an (Feynman) amplitude to any graph, and correlators are their infinite sums.\\

In most of the cases (for interacting theories), these Feynman amplitudes are divergent quantities and become finite after renormalization. Among the renormalization techniques, those of Bogoliubov-Parasiuk-Hepp-Zimmermann (BPHZ) consists in eliminating divergences by adding counter-terms to amplitudes; the counter-terms are constructed in recursive way in the presence of subdivergent amplitudes: it is the Bogoliubov-Parasiuk (BP) preparation.\\

Later, Dirk Kreimer have shown that the BP preparation allows to define a Hopf structure on some set of Feynman diagrams. The co-product is done by the decomposition of diagrams in their divergent subdiagrams, whereas antipode exists due to connected grading on diagrams. Most important is the fact that counter-terms are given by some deformation of antipode.\\

More later, Alain Connes and Dirk Kreimer introduced a more natural picture for understanding \emph{in the same time} the relations which give counter-term and renormalized amplitude \footnote{The above deformation of antipode is more natural in this scheme.}: it is the Birkhoff decomposition. Roughly speaking, counter-term and renormalized amplitude are the duals (in the sense of Gelfand-Naimark) of the Birkhoff decomposition components of a loop \cite{acon02}.\\

An important physical concept which can be derived from this approach is the renormalization group. In physical theories, the renormalization group describes the evolution of physical parameter with the energy scale; it is described by one-parameter subgroup of the group of diffeomorphisms in the space of physical parameters, whose generator is called the \emph{beta function}. In Connes-Kreimer theory, the renromalization group is a subgroup of the structure space of the Hopf algebra of graphs.

\section{Feynman diagrams}

In this section, we will recall the uses of Feynman diagrams in the perturbative calculus of correlation functions \cite{paet}. We will only examinate the case where the fields space is a finite dimensional vector space. Let us consider a finite dimensional vector space V with a Lebesgue measure dx and define:
\begin{eqnarray*}
<\phi> := \int_{V}dx \phi(x)\cdot e^{-S[x]},
\end{eqnarray*}
where S is the action function $\phi$ is a polynomial function on V, i.e. 
\begin{eqnarray*}
\phi := \sum_{\alpha}f_{1}^{\alpha}\cdot...\cdot f_{N(\alpha)}^{\alpha},\quad for \quad f_{i}^{\alpha}\in V^{*}.
\end{eqnarray*}
$<\phi>$ is called the \emph{correlator} of $\phi$.\footnote{In a realistic QFT, the correlator is $<\phi> := \int_{V}dx \phi(x)\cdot e^{-i\hbar S[x]}$, where V is an infinite dimensional vector space.} 
This is linear in $\phi$ so it suffices to calculate: 
\begin{eqnarray*}
<f_{1}\cdot...\cdot f_{N}> = \int_{V}dx f_{1}\cdot...\cdot f_{N}(x)\cdot e^{-S[x]},
\end{eqnarray*}
When $\phi$ = 1, a monomial of degree 0, we note Z := $<1>$ and call it the \emph{partition function}. 

\subsection{Free action correlators}
It is the case where the action is free; for our purpose, it means that the action is a definite positive quadratic form B/2. The main result is that Z is finite and can be normalized (from the Lebesgue measure) to one.
So, for V = \textbf{R} with the standard inner product and the standard measure dx, one have the normalized measure $\tilde{dx} := dx/\sqrt{\pi}$.
For $V = \textbf{R}^d$ with the inner product B/2 and the standard measure dx$^d$, one have the normalized measure $\tilde{dx}^d := dx^d/(det(B/2\pi))^{-1/2}$. 

\subsection{Wick theorem}
By normalizing the partition function, it permits to obtain the free action correlators from 'contractions', i.e.\\
for even N:
\begin{eqnarray}
<f_{1}\cdot...\cdot f_{N}>_{free} = \sum_{s\in S_{N}/pair}B^{-1}(f_{s(1)},f_{s(2)})\cdot ...\cdot B^{-1}(f_{s(N-1)},f_{s(N)}),
\label{wt}
\end{eqnarray}
where s , s'$\in S_N$, are identified when they define the same term of (\ref{wt}) (these terms are called Wick pairings);\\ 
for odd N, it is zero.\\
\\
\textsl{Proof}:
The proof is subdivided in two parts and follows closely \cite{paet}.\\
\\
\textbf{1. Reduction to dimension one.} 
As it stand, both sides of (\ref{wt}) are symmetric linear forms on $(V^*)^{\otimes N}$, so it suffices to verify the equality for a symmetric element of $(V^*)^{\otimes N}$.
Further, any positive definite inner product can be diagonalized in many orthonormal basis, and taking f(x) = $\alpha$, we see that the integral is a sum of 1D integrals of the form $K(l_1,...,l_d)<x_{1}^{l_1}> ... <x_{i}^{l_d}>$ where d is the dimension of the vector space, $\sum_{j=1}^d l_j = N$, and $<x_i^{l_i}>$ are correlators calculated in \textbf{R} with its standard inner product and its normalized measure.\\
\\ 
\textbf{2. Calculation of $<x^{N}>$.} This will be calculated from the linear perturbed action $\frac{x^{2}}{2} + J\cdot x$.
Define the perturbed partition function
\begin{eqnarray*}
Z(J) := \int_{\textbf{R}} \frac{dx}{\sqrt{2\pi}} e^{-\frac{x^2}{2} +J\cdot x} ,
\end{eqnarray*}
so the free action correlator for $x^{N}$ is:
\begin{eqnarray*}
<x^N>_{free} &=& \int_{\textbf{R}}\frac{dx}{\sqrt{2\pi}}x^{N} \cdot e^{-\frac{x^2}{2}} \\
&=& (\partial_J)^{N}Z(J)|_{J=0}. 
\end{eqnarray*}
Now, we will check that:\\
1) Z(J) = $e^{J^{2}/2}$ ;\\
2) $<x^{N}>_{free}$ vanishes for odd powers and for even N = 2p, 
\begin{eqnarray*}
<x^{N}>_{free} = \frac{N!}{2^{p}\cdot p!}.
\end{eqnarray*}
For 1), we complete the linear perturbed action to a square:
\begin{eqnarray*}
Z(J) = \int_{\textbf{R}} \frac{dx}{\sqrt{2\pi}} e^{\frac{-x^2 + J^2}{2}} = e^{J^2/2}.
\end{eqnarray*}
For 2), $e^{J^2/2} = \sum_{i=0}^\infty \frac{J^{2i}}{2^{i}\cdot i!}$,
so $(\partial_J)^{N}Z(J)|_{J=0}$ is zero for odd N, \\
and for N = 2p, this is $N!$ times the coefficient of $J^{N}$ in $e^{J^2/2}$, i.e.
\begin{eqnarray*}
 (\partial_J)^{N}Z(J)|_{J=0} = N!\cdot \frac{1}{2^p \cdot p!} \quad \textrm{for } N = 2p.
\end{eqnarray*}
\textbf{Remarks}:\\
From the definition, we can write:
\begin{eqnarray*}
	<x^{N}> = <Id \underbrace{...}_{N times}Id> ,
\end{eqnarray*}
where Id is the identity form on V.
On the other hand, noticing that the inverse form of Id by Id is 1, one have:
\begin{eqnarray*}
	\sum_{s \in S_N/pair}1 &=& |S_N/pair|\\ 
	&=& \frac{\bigl(\frac{N(N-1)}{2}\bigr)\cdot...\cdot \bigl(\frac{2.1}{2}\bigr)}{p!}\\
	&=& N!\cdot \frac{1}{2^p \cdot p!}. 
\end{eqnarray*}
so $<x^N>$ verifies the Wick theorem.

\subsection{Perturbations of free actions.}
Now, we consider the general case where the action is not free and is 'infinitesimaly' perturbed ,i.e. of the form:
\begin{eqnarray*}
	S[x] = S_{free}[x] - Q(x) \quad for \quad Q(x) = \sum_{m\geq 1} \frac{g_m}{m!}Q_m(x^{\otimes m})
\end{eqnarray*}
where:
\begin{itemize}
	\item $Q_m$ is a symmetric m-form on V, i.e. :\\
	$Q_m = \omega_1 \otimes ... \otimes \omega_m = \omega_{s(1)}\otimes ... \otimes \omega_{s(m)} \quad \forall s \in S_{m} , \omega_i \in V^* $
	\item $g_m$'s are formal parameters used to separate the contributions from $Q_m$'s, 
\end{itemize}
\subsection{Feynman graphs}
Now, we will define Feynman graphs in order to use them in the calculation.
A \textsl{graph} is a collection of vertices and edges; each edge 'contracts' two vertices and the number of edges incident on a vertex is called its \textsl{valence}.\\
A \textsl{Feynman graph} is a graph with:\\
(a) 'external' (or labeled) vertices, which are all univalent.\\
(b) 'internal' (or bar-labeled) vertices, of different valences.\\
We denote by G(N,$\vec{n}$) the set of equivalence classes of all graphs which have N external vertices, and $n_m$ m-valent internal vertices , $m \geq 1 , n_m \in \vec{n}$.\\ 
In order to calculate correlators, we associate to every Feynman graph $\Gamma$ a  \textsl{Feynman amplitude} $F_\Gamma $ following the so-called Feynman rules:\\
(1) Place at each vertex some symmetric tensor in V;
\begin{itemize}
	\item At external vertex j place form $f_j$,
	\item At each m-valent internal vertex place tensor $Q_m$  
\end{itemize}
(2) And contract the tensors along the edges of the graph using the form $B^{-1}$.\\
\\
\textbf{Example.} Consider:
\begin{eqnarray*}
	\Gamma_1 = 1 \longleftrightarrow \bar{4} \longleftrightarrow
\bar{4} \longleftrightarrow 2
\end{eqnarray*}
then:
\begin{eqnarray*}
F_{\Gamma_1} &:=& B^{-1}(f_1 \otimes Q_4,Q_4 \otimes f_2),\\
&:=& B^{-1}(f_1,\omega_1)(B^{-1}\omega_1,\omega_2,)B^{-1}(\omega_2,\omega_3)
B^{-1}(\omega_3,\omega_4)B^{-1}(\omega_4,f_2)
\end{eqnarray*}
\\
$F_{\Gamma_1}$ can be expressed as product of $B_1$ terms, and it is not difficult to see that it is a term in the correlator $<f_1 \cdot f_2 \cdot (Q_4)^2>_{free}$.\\
More precisely, if we define the graph 
\begin{eqnarray*}
  \Gamma_2 := 1 \longleftrightarrow 2 \quad \cup \quad \bar{4} \longleftrightarrow \bar{4},	
\end{eqnarray*}
we have:
\begin{eqnarray*}
<f_1 \cdot f_2 \cdot Q_4 \cdot Q_4>_{free} = \sum_{i=1}^2|Aut(\Gamma_i)|^{-1}2!(4!)^2F_{\Gamma_i}.
\label{ex1}
\end{eqnarray*}
This relation is due to :
\begin{itemize}
	\item the symmetry of $Q_4$, i.e. each pairing is equal to $F_{\Gamma_i}$ , i = 1,2; 
	\item the fact that the factor $|Aut(\Gamma_i)|^{-1}(4!)^22!$ is the number of pairings of each i-term needed to obtain the correlator:
   \begin{itemize}
	  \item $(4!)^2$ gives terms generated by the symmetric tensor $Q_4$;
	  \item $2!$ provides terms from the discernability of the two $Q_4$ by the Wick formula;  
	  \item $|Aut(\Gamma_i)|^{-1}$ restricts permutation to pairings; 
    \end{itemize}
	\end{itemize}
How to calculate $|Aut(\Gamma)|$ for a graph $\Gamma$?\\
From relation (\ref{ex1}), we deduce that $|Aut(\Gamma)|$ cures permutations from 1) $|Q_E(\Gamma)| := \prod_{m}(m!)^{n_m}$ and 2) $|Q_V(\Gamma)| := \prod_{m}n_m!$.\\
1) and 2) are relative with permutations of internal 1/2-edges, and internal vertices, respectively. So it is more convenient to write $|Aut(\Gamma)|$ as:
\begin{eqnarray*}
	|Aut_E(\Gamma)|\cdot |Aut_V(\Gamma)|
\end{eqnarray*}
where $|Aut_E(\Gamma)|$ is the number of edges permutations which preserve incidence on vertex (this construction leads to consider Feynman graph as ribbon graph, i.e. graph with cyclic ordering on the incident 1/2-edges of each vertex),\\
and $|Aut_V(\Gamma)|$ a 'by hand' number (it is always 1 for nonvacuum graph).  
Now, we can derive the more general formula : the Feynman's theorem.
\subsection{Feynman's theorem}
\textsl{Theorem.} The correlator of monomial for the perturbated action
\begin{eqnarray*}
	S[x]= S[x]_{free} - Q[x]
\end{eqnarray*}
is a formal power series \\
\begin{eqnarray*}
	<f_1 \cdot ... \cdot f_N> = \sum_{\vec{n}}\prod {g_i^{n_i}}\sum_{\Gamma \in FG(N,\vec{n})}\frac{F_\Gamma}{|Aut(\Gamma)|}
\end{eqnarray*}
where Aut($\Gamma$) is the group of automorphisms of $\Gamma$ which fix the external vertices.\\
\\
\textit{Proof}: First, it can be reduced to the calculation of free correlators.\\
Indeed, we have:
\begin{eqnarray*}
	<f_1 \cdot ... \cdot f_N> &=& <f_1 \cdot ... \cdot f_N \cdot e^Q>_{free}\\
	&=& <f_1 \cdot ... \cdot f_N \prod_{m\geq 1} \sum_{n_m\geq 0} \frac{1}{n_m!} 
	\Bigl(\frac{g_m Q_m}{m!}\Bigr)^{n_m}>\\
	&=& <f_1 \cdot ... \cdot f_N \sum_{\vec{n}} \prod_{m\geq 1} \frac{1}{n_m!}
	\Bigl(\frac{g_m Q_m}{m!}\Bigr)^{n_m}>\\
	&=& \sum_{\vec{n}} \prod {g_i^{n_i}} \cdot \frac{<f_1 ...f_N \prod_{m\geq 1} (Q_m)^{n_m}>}{\prod_{m\geq 1}n_m!(m!)^{n_m}}   
\end{eqnarray*}
\textsl{Remarks:}\\ 
1) We have used the Leibniz product: 
\begin{eqnarray*}
	\prod_{m=1}^v \sum_{i=0}^r (a_m)^i = \sum_{\vec{i}} \prod_m (a_m)^{i_m}, \quad with \quad \vec{i} = (i_1,...,i_v) \quad and \quad \sum_{m=0}^v i_m \leq vr. 
\end{eqnarray*}
2) So the indice m of $\vec{n}_m$ is only superfluous when we do the polynomial expansion of $e^Q$.\\
Next, we can think of each m-valent vertex as m neighbouring 1-valent vertices; then, any graph $\Gamma \in G(N,\vec{n})$ defines as many as $|Aut(\Gamma)|^{-1}\prod_{m\geq 1}n_m! m!^{n_m}$ different pairings of such 1-valent vertices (\cite{edwi96}, Witten). Q.E.D \\
Next, we will see that Hopf algebra structure can explained easily Feynman's calculus.
 
\section{Hopf algebra on Feynman diagrams}

Due to Gelfand-Naimark duality, a group structure on Hausdorff space generates an algebraic structure called co-group on the algebra of coordinates, then becoming a Hopf algebra. One can also find Hopf structures in combinatorial problems, where the co-product gives the decomposition mode of element in its components.
In renormalization theory, the BP preparation, which is a combinatorial formula, is a deformation of the antipode of a Hopf structure on Feynman diagrams having as co-product the decomposition in subdivergences.

\subsection{Hopf algebra}
We recall that a Hopf algebra $\calh$ is an algebra with a co-group structure ($\Delta$: $\calh$ $\rightarrow$ $\calh\otimes \calh$, $\epsilon$: $\calh$ $\rightarrow$ \textbf{R}, S: $\calh$ $\rightarrow$ $\calh$), compatible with the algebra structure on $\calh$, i.e. 
\begin{enumerate}
	\item Co-associativity: ($\Delta\otimes id)\Delta = (id\otimes \Delta)\Delta$;
	\item Co-unit: ($\epsilon \otimes id)\Delta = (id\otimes \epsilon)\Delta$ = id; 
	\item Co-inverse: m(S$\otimes$id)$\Delta$ = m(id$\otimes$S)$\Delta$ = u$\epsilon$ ,\\
where m and u are the multiplication and the unit respectively. $\Delta$, $\epsilon$, and S are called the \emph{co-product}, the \emph{co-unit} and the \emph{antipode} (or co-inverse) respectively. 
\end{enumerate}
On can define the antipode from an operation on the morphism set Mor($\calh$,A) for A(m$_A$,$\eta$) an unital algebra: it is the convolution, denoted by $\ast$, which is dual to the pointwise product on Mor(Struct(A),Struc($\calh$)), i.e.
\begin{eqnarray}
	f\ast g = m_A(f\otimes g)\Delta , \quad f,g\in \textrm{Mor($\calh$,A)}
\end{eqnarray}
We have:
\begin{eqnarray*}
	m_A(f\otimes \eta \epsilon)\Delta(a) &=& m_A(f\otimes \eta \epsilon)(a'_i\otimes a_i) , \quad \textrm{for} \quad \Delta(a) = a'_i\otimes a_i,\quad a\in \calh\\
	&=& f(a'_i)\epsilon(a_i),\\
	&=& f(a'_i)\epsilon(a_i)) \quad \textrm{(linearity of f)},\\
  &=& f((id\otimes \epsilon)\Delta(a)),\\
  &=& f(a) \quad \textrm{(co-unit)}.
\end{eqnarray*}
and, 
\begin{eqnarray*}
	m_A(\eta \epsilon \otimes f)\Delta(a) = f(a).  
\end{eqnarray*}
The last two equalities show that $\eta \epsilon$ is the neutral element of the convolution; consequently, the definition of co-inverse means that the antipode is the inverse of the identity function on $\calh$ by the convolution, and then is unique.\\

Another structure which can give the antipode on a bi-algebra H ( a 'Hopf algebra' without antipode) is a connected grading on H, compatible with the bi-algebra structure, i.e.   
\begin{eqnarray*}
&& \textrm{Grading}: H = \oplus_{n=0}^{\infty}H^{(n)},\\
&& \textrm{Connectedness}: H^{(0)} = \textrm{u}(\textbf{R}),\\
&& \textrm{Compatibility}:	m(H^{(n)}\otimes H^{(m)}) \subseteq H^{(n+m)} \quad \textrm{et} \quad \Delta(H^{(n)}) \subseteq \oplus _{p+q = n} H^{(p)}\otimes H^{(q)}.
\end{eqnarray*}
Leaving aside this existence of the antipode, the connected grading implies a co-product of the form:
\begin{eqnarray}
	\Delta(a) =  a\otimes \textbf{1} + \textbf{1}\otimes a + a'_j\otimes a_j, \quad \textrm{for deg(a) = n}
\end{eqnarray}
with deg($a'_i$) and deg($a_i$) between 1 and n-1 (*).
Indeed, we have \cite{doma02}:
\begin{eqnarray*}
	\epsilon(ab) &=& \epsilon(a)\epsilon(b) \quad \forall a,b\in A,	\textrm{ so } \epsilon(\textbf{1}) = 1,\\
	\textrm{and } \Delta a &=& \alpha a\otimes \textbf{1} + \beta \textbf{1}\otimes a + a'_j\otimes a_j, \quad \alpha,\beta \in \textbf{C},\\
	\textrm{so}\quad a &=& (\epsilon\otimes id)\Delta a = \alpha \epsilon(a) + \beta a + \epsilon(a'_j)a_j, \\
	&=& (id\otimes \epsilon)\Delta a = \alpha a + \beta \epsilon(a) + a'_j\epsilon(a_j).
	\end{eqnarray*}
As deg(a'$_j$) $\neq$ deg(a), deg(a$_j$) $\neq$ deg(a), then:
\begin{eqnarray*} 
	\left\{ 
\begin{aligned}
  &\epsilon(a) = 0 \quad \textrm{and } \alpha =\beta = 1, \quad \forall a\in A, deg(a)\geq 1,\\
  \textrm{and } &(\alpha + \beta) = 1 ,\quad \forall a\in A, deg(a) = 0. 
\end{aligned}
  \right.
\end{eqnarray*}
And the antipode is given by the geometric series:
\begin{eqnarray*}
	S = \sum_{i=0}^{\infty}u_i, \quad \textrm{with} \quad u_{i+1} = (\eta \epsilon - id)\ast u_i, \quad u_0 = \eta \epsilon.
\end{eqnarray*}
Indeed, the partial sum $S_i = \sum_{l=0}^i u_l$ satisfies the following relation: 
\begin{eqnarray}
	u_{i+1} + S_i = (\eta \epsilon - id)\ast S_i + u_0 , \quad \textrm{i.e.} \quad id\ast S_i = u_0 - u_{i+1}. 
\end{eqnarray}
In the other hand, 
\begin{prop} \cite{hijo04} A connected grading on A implies u$_k$(a) = 0 $\forall$ k $>$ deg(a), a$\in$A, so u$_{\infty}$ = 0.
\label{gr}
\end{prop}
\textbf{Proof}:\\
For a = $\gamma$\textbf{1}, we have: 
\begin{eqnarray*}
	u_1= (\eta \epsilon - id)(\gamma\textbf{1}) = 0, \textrm{ and } u_1*u_i(a) &=& m(u_1\otimes u_i)\Delta(\gamma\textbf{1}), \quad \forall i\geq 0,\\
	&=& m(u_1\otimes u_i)(\gamma \textbf{1}\otimes \textbf{1}) = 0.
\end{eqnarray*}
Suppose the claims holds for all b$\in$A $|$ deg(b)$\leq$ n-1, then 
\begin{eqnarray*}
	u_k(a) &=& u_1*u_{k-1}(a) = m(u_1\otimes u_{k-1})\Delta(a) , \quad k > deg(a) = n,\\
	&=& m(u_1\otimes u_{k-1})(a\otimes \textbf{1} + \textbf{1}\otimes a + a'_j\otimes a_j),\\
	&=& 0 ,\quad (u_1(\textbf{1}) = 0, \quad k-1 > deg(a) -1 \geq deg(a'_j), \quad k-1 > deg(a) -1 \geq deg(a_j)).\\
\textrm{Q.E.D}.
\end{eqnarray*}
Then, we have:
\begin{eqnarray*}
	S(a) &=& u_0(a) + u_1(a) + \sum_{i=2}^{n}u_{i}(a), \quad \textrm{for deg(a) = n}\\
	&=& - a + \sum_{i=1}^{n-1}u_{i}\ast u_1(a), \quad \textrm{($\epsilon$(a) = 0)}\\
	&=& - a + m(\sum_{i=1}^{n-1}u_{i}\otimes u_1)\Delta(a),\\
	&=& - a + m(\sum_{i=1}^{n-1}u_{i}\otimes u_1)(a\otimes 1 + 1\otimes a + a'_j\otimes a_j) , \\
	&=& - a + \sum_{i=1}^{n-1}u_{i}(a'_j) u_1(a_j), \quad \textrm{deg(1) = 0}.
\end{eqnarray*}
Finally, from Proposition \ref{gr}, we obtain the folllowing recursive formula:
\begin{eqnarray}
	S(a) = - a + S(a'_j)a_j, \quad \textrm{with} \quad \Delta(a) = a\otimes 1 + 1\otimes a + a'_j\otimes a_j.
	\label{at}
\end{eqnarray}
This kind of formula (\ref{at}) is present in combinatorial formulas of perturbative renormalization.

\subsection{Bogoliubov-Parasiuk preparation and Hopf structure}

In BPHZ renormalization, when the divergent amplitude does not contain any divergent subgraph, it suffices to add it a counter-term to obtain a finite (or renormalized) amplitude. In the case with subdivergences, one substract each subdivergence and after add to the resulting (or prepared) amplitude (with no subdivergence) the corresponding counter-term: it is the BP preparation.\\

Most precisely, for a divergent connected graph $\Gamma$,\\
the prepared graph P($\Gamma$) is given by:
\begin{eqnarray*}
	P(\Gamma) = F(\Gamma) + \sum_{\gamma \subset \Gamma}C(\gamma)F(\Gamma/\gamma),
\end{eqnarray*}
where the sum is on all subdivergent graphs, (F is the Feynman amplitude map)\\
the counter-term by:
\begin{eqnarray}
	C(\Gamma) = -T(P(\Gamma)) = -T(F(\Gamma) + \sum_{\gamma \subset \Gamma}C(\gamma)F(\Gamma/\gamma)),
	\label{ct}
\end{eqnarray}
and the renormalized amplitude by:
\begin{eqnarray}
	R(\Gamma) = P(\Gamma) + C(\Gamma),
	\label{reap}
\end{eqnarray}
where T is the operation defining the counter-term for a graph without subdivergence.
Now, let's construct a Hopf structure on Feynman diagrams with antipode in the form (\ref{ct}).

Consider the commutative real algebra $\calh$ generated by all unions of one-particle irreducible (or 1PI) Feynman graphs, with union as product.\\
Define on the basis \footnote{To define an algbra morphism, it suffices to do it on a basis, due to the definition of basis and of morphism.} the co-product by: 
\begin{eqnarray*}
	\Delta(\Gamma) &=& \sum_{\gamma \subseteq \Gamma}\gamma \otimes \Gamma /\gamma ,\\
	&=& \Gamma \otimes 1 + 1 \otimes \Gamma + \sum_{\gamma \subset \Gamma}\gamma \otimes \Gamma /\gamma , 
\end{eqnarray*}
where the sum is on all subdivergent graphs.\\
Later, the second equality will be useful for the construction of the antipode.\\
The next structure is the connected grading, that one can define, for a basis element $\Gamma$, by the number
	\[\nu(\Gamma) = v(\Gamma) - 1 , \quad \textrm{where v($\Gamma$) is the number of internal vertices of $\Gamma$}.
\]
and compatible with the product on $\calh$.\\
Another grading, important for the construction of the renormaliztion group, is giving by the loop number L defined on generator diagrams by
\begin{eqnarray*}
	L(\Gamma) = I(\Gamma) - \nu(\Gamma), \quad \textrm{where I($\Gamma$) is the number of internal lines of }\Gamma.  
\end{eqnarray*}
From the previous section, there exists an antipode S on $\calh$, given by:
\begin{eqnarray}
	\textrm{S}(\Gamma) = - \Gamma + \sum_{\gamma \subset \Gamma}S(\gamma)\Gamma/\gamma.
	\label{ath}
\end{eqnarray}
One can see that the counter-term is identical to the antipode (\ref{ath}) when T is the identity.
\begin{note}
In fact, there is many formulas giving counter-terms as those of Zimmermann, or those of Dyson-Salam, but they define all \emph{the} same antipode (\ref{ath}), \cite{hijo04}.
\end{note}
Another relation between R and F is given by the convolution:
\begin{eqnarray*}
	C\ast F(\Gamma) &=& m_A(C\otimes F)\Delta(\Gamma),\\
	&=& m_A(C\otimes F)(\Gamma \otimes 1 + 1 \otimes \Gamma + \sum_{\gamma \subset \Gamma}\gamma \otimes \Gamma /\gamma),\\
	&=& F(\Gamma) + C(\Gamma) + \sum_{\gamma \subset \Gamma}C(\gamma) F(\Gamma /\gamma),\quad \textrm{(Feynman rules)}
\end{eqnarray*}
i.e.
\begin{eqnarray}
	C\ast F = R.
	\label{bpf}
\end{eqnarray}
 
\section{Pertubative renormalization and (Birkhoff) decomposition}

We will show that, in the dimensional regularization-minimal substraction (DimReg-MS) renormalization, the BP formula (\ref{bpf}) is dual to the Birkhoff decomposition of a loop on the Riemann sphere 
\textbf{\^C}:= C$\cup \left\{\infty\right\}$ with value in G := Struc($\calh$), the structure space of $\calh$. 
Recall that the Birkhoff decomposition of a loop $\gamma: C \rightarrow G$, where C is a curve on \textbf{\^C}, is given by:  
\begin{eqnarray}
	\gamma(z) = \gamma_-(z)^{-1}\gamma_+(z), \quad z\in C
\end{eqnarray}
$\gamma_\pm$ are boundary values of holomorphic maps 
$\gamma_{h,\pm}: C_{\pm} \rightarrow G$, with C$_-$ the connected component of the complement of C containing $\infty$ and C$_+$ the bounded component. Then
when $\gamma$ can be extented into holomorphic map $\tilde{\gamma}$ on C$_+$; one can take: $\gamma_{h,+}$ = $\tilde{\gamma}$ and $\gamma_-$ = 1.   

The principal result is that Mor(\textbf{\^C},G) equipped with pointwise product have as dual the set Mor(C(G)=$\calh$, C(\textbf{\^C})) equipped with the convolution. Indeed, we have: 
\begin{eqnarray*}
	m_G(\gamma_1\otimes \gamma_2)\tau: z \mapsto \gamma_1(z)\gamma_2(z) , \quad \tau: z \mapsto z\otimes z,
\end{eqnarray*}
and
\begin{eqnarray*}
	C(m_G(\gamma_1\otimes \gamma_2)\tau) &=& C(\tau)(C(\gamma_1)\otimes C(\gamma_2))C(m_G),\\
	&=& m_{C(\textbf{\^C}))}(C(\gamma_1)\otimes C(\gamma_2))\Delta,\\
	&=& C(\gamma_1)\ast C(\gamma_2).
\end{eqnarray*}

By definition of Feynman rules, the Feynman amplitude map preserves the product on $\calh$, then defines an element of Mor($\calh$,A) for a commutative algebra A with unit, and in this way an element of Map(Struc(A),G) , Struc := C$^{-1}$.
So, from (\ref{bpf}), we have:
\begin{eqnarray*}
	Struc(R) &=& Struc(C\ast F), \\
	&=& Struc(C)\cdot Struc(F), \quad \textrm{where $\cdot$ is the pointwise product on Mor(Struc(A),G)},
\end{eqnarray*}
i.e.
\begin{eqnarray}
	Struc(F) = Struc(C)^{-1}\cdot Struc(R).
	\label{ckde}
\end{eqnarray}
In renormalization theory, the divergence or not of the amplitude F depends on its value, so by definition, the set A$_+$\footnote{A$_+$ is in fact a subalgebra of A.} of 'finite' elements of A is composed by values of renormalized amplitudes \footnote{The operation T of BP can be defined as being the projection of P($\calh$) on P($\calh$)$\backslash$P($\calh$)$\cap$A$_+$.}.\\

For example, in a D-dimensional QFT, when we use the DimReg-MS renormalization, we have:\\
A is the ring of meromorphic functions in the neighborhood of z = D, A$_-$ the subring of polynomial functions in (z-D)$^{-1}$, A$_+$ the subring of regular functions in D, and the equality (\ref{ckde}) is the Birkhoff decomposition of the loop Struc(F). The loop $\gamma$ have $\gamma_+(D)$ as renormalized value in D \cite{acdk99}. 

\section{The renormalization group}

In this section, we will derive the renormalization group of the $\phi^3$-theory.\\
The renormalization group comes from the idea that (renormalized) physical parameters depend on the energy scale $\mu$. In the DimReg-MS scheme, the scale dependence of Feynman amplitude is introduced by the restauration of the amplitude dimension after the extension D$\mapsto$ D+$\epsilon$ =: z, $\epsilon\in$ \textbf{C}$\backslash\left\{0\right\}$. 

We will use the following propriety: the singular part of the Birkhoff decomposition is scale-independent \cite{acdk00}. Because of this, one can construct an one-paramater subgroup \textbf{rg}$_t$ of group G := Struc($\calh$) whose generator is the beta function of renormalized loops \cite{acdk99,acdk00}. By definition, renormalized loops are integral curves of the beta vector field.\\

Due to the linearity with rapport to graphs and the homogeneity with rapport to $\mu$ of amplitudes, the effect of scale variation $\mu \mapsto e^t\mu$ on amplitudes can be induced by the action on $\calh$ defined by:
\begin{eqnarray*}
	\theta_t(\Gamma) = e^{tL(\Gamma)}\Gamma \quad \forall \Gamma \in \calh.
\end{eqnarray*}
(The presence of the grading is necessary for $\theta$ to be a morphism) and we have \cite{acdk00}:
\begin{eqnarray}
	\gamma_{e^t\mu}(z) = \theta_{t(D-z)}(\gamma_{\mu}(z)) \quad \forall t\in \textbf{R} \quad \forall D-z\in\textbf{C}\backslash\left\{0\right\}, 
\end{eqnarray}
where $\gamma$ := Struc(F). \\
So we obtain the one-parameter subgroup Struc($\theta_t$) of Diff(G).
To study a flow of the form $\gamma_{\mu}^+$(D), $\mu\in$\textbf{R}, one can consider the one-parameter subgroup:
\begin{eqnarray}
	\textbf{rg}_t = lim_{z\rightarrow D} \gamma_-(z)\theta_{t(D-z)}(\gamma_-(z)^{-1}).
\end{eqnarray}
The limit exists because of the holomorphicity of $\gamma_{\mu +}$ in z = D, whereas the scale-independence of $\gamma_{\mu -}$ induces the group structure on \textbf{rg}$_t$, and we have:
\begin{eqnarray*}
	\gamma_{e^t\mu}^+(z) &=& \theta_{t(D-z)}(\gamma_{\mu-}(z))\theta_{t(D-z)}(\gamma_{\mu}(z)),\\
	&=& \gamma_{\mu-}(z)\theta_{t(D-z)}(\gamma_{\mu}(z)), \quad (\textrm{scale independence of } \gamma_-),\\
	&=& \gamma_{\mu-}(z)\theta_{t(D-z)}(\gamma_{\mu-}(z)^{-1}\gamma_{\mu}^+(z))\\
	&=& \gamma_{\mu-}(z)\theta_{t(D-z)}(\gamma_{\mu-}(z)^{-1})\theta_{t(D-z)}(\gamma_{\mu}^+(z)). 
\end{eqnarray*}
A the limit z=D, we have:
\begin{eqnarray}
	\gamma_{e^t\mu}^+(D) = \textbf{rg}_t \theta_{0}(\gamma_{\mu}^+(D)),
\end{eqnarray}
then:
\begin{eqnarray*}
	\beta = (\gamma^+_{e^{(.)}}(D))_*(\frac{\partial}{\partial \mu}) = (\gamma^+_{(.)}(D))_*(e^{(.)})_*(\frac{\partial}{\partial \mu}) = (\gamma^+_{(.)}(D))_*(\mu \frac{\partial}{\partial \mu}).
\end{eqnarray*}
where $\partial/\partial \mu$ is the vector field on \textbf{R} defined by: ($\partial/\partial \mu$)(Id) := 1, and $\beta$ the generator of \textbf{rg}$_t$.\\
The first equality means that $\gamma^+_{e^{(.)}}(D)$ is an integral curve of $\beta$. Here, as $\beta$ is the generator of the subgroup \textbf{rg}$_t$, then its flows are \textbf{rg}$_t$-valued.

\section{Conclusion}
The Connes-Kreimer approach on renormalization theory allows to consider renormalized amplitudes and their counter-terms as being the terms of Birkhoff decomposition: in dimensional regularization, the BP preparation is dual to the Birkhoff decomposition. Moreover, the renormalization group is naturally expressed by a subgroup of the structure space of Feynman diagrams.\\
However, the CK theory does not answer the following question:
Can we explain more the scale dependence of physical theories?
In our next paper, we will give a model which exhibits more \emph{naturally} this propriety.


\begin{thebibliography}{20}
\bibitem{paet}Ivan , \emph{QFT Seminar notes 2. Feynman calculus}, from Pavel Etingof's talk.
\bibitem{edwi96} Edward Witten, \emph{Perturbative renormalization I.}, QFT porgram at IAS, \textbf{1996}.
\bibitem{acdk99} Alain Connes, Dirk Kreimer, \emph{Renormalization in Quantum Field Theory and the Riemann-Hilbert problem I: the Hopf algebra structure of graphs and the main theorem}, preprint: hep-th/9912092, \textbf{1999}.
\bibitem{acdk00} Alain Connes, Dirk Kreimer, \emph{Renormalization in Quantum Field Theory and the Riemann-Hilbert problem II: the $\beta$-function, diffeomorphisms and the renormalization group}, preprint: hep-th/0003188, \textbf{2000}.
\bibitem{acon02} Alain Connes, \emph{Sym\'etries galoisiennes et renormalisation}, preprint: QA/0211199, \textbf{2002}.
\bibitem{acmm05} Alain Connes, Matilde Marcolli, \emph{Quantum fields and motives}, preprint: hep-th/0504085, \textbf{2005}.
\bibitem{hijo04} H\'ector Figueroa and Jos\'e F. Gracia-Bondia, \emph{The uses of Connes and Kreimer's algebraic formulation of renormalization theory}, preprint: hep-th/0301015, \textbf{2004}.
\bibitem{doma02} Dominique Manchon, \emph{Hopf algebras, from basics to applications to renormalizations}, Extended version of lectures given at Bogota, \textbf{2002}.
\end{thebibliography}
\end{document}